\documentclass{article}
\begin{document}

\def\d{\partial}
\def\R{{\bf R}}
\def\f#1#2{\frac{#1}{#2}}
\def\be{\begin{equation}}
\def\ee{\end{equation}}
\def\bea{\begin{eqnarray}}
\def\eea{\end{eqnarray}}

\date{}

\title{Fuchsian analysis of singularities in Gowdy spacetimes beyond 
analyticity}
\author{Alan D. Rendall \\
Max--Planck--Institut f\"ur Gravitationsphysik \\
Am M\"uhlenberg 1, 14476 Golm, Germany}

\maketitle

\begin{abstract} 
Fuchsian equations provide a way of constructing large classes of spacetimes 
whose singularities can be described in detail. In some of the applications
of this technique only the analytic case could be handled up to now. This 
paper develops a method of removing the undesirable hypothesis of analyticity.
This is applied to the specific case of the Gowdy spacetimes in order to
show that analogues of the results known in the analytic case hold in the
smooth case. As far as possible the likely strengths and weaknesses of the 
method as applied to more general problems are displayed.
\end{abstract}   

\section{Introduction}

The theory of Fuchsian equations has been applied to analyse singularities
in a variety of classes of spacetimes in general relativity. The term
\lq Fuchsian equations\rq\ has not always been used in the literature
on this subject and in this paper it denotes a certain class of singular
differential equations in a generic way. The existing results in this 
area will be surveyed briefly below. In this approach spacetimes containing 
singularities are parametrized by some functions which play the role of 
data on the singularity. In some cases it was necessary to assume the 
analyticity of these functions. In other cases smoothness was sufficient.
The aim of this paper is to develop ways of removing the analyticity
requirement. These will be illustrated by the case of Gowdy spacetimes
which represent an ideal laboratory for testing new ideas on the 
mathematical treatment of the Einstein equations.

It may not be immediately obvious why the apparently technical distinction
between analytic ($C^{\omega}$) functions and smooth ($C^\infty$) functions
should be significant with a view to physical applications. There are at 
least two reasons why it is important. The first is that the physical notion 
of causality cannot be reasonably formulated within the class of analytic
functions, since the unique continuation property of the latter means
that the solution of an equation at one point influences its behaviour
at all other points, and not only at causally related points. Connected
with this is the fact that solutions of an equation do not depend 
continuously on initial data in any useful sense. For more discussion
of these points see \cite{friedrich00}, particularly section 2.4. The
second, which is also a direct consequence of the unique continuation
property of analytic functions, is that there is not the same freedom
to construct solutions with certain interesting properties within the
analytic class. An example of this will be given in section 
\ref{discussion} below. 

There are a number of results on Fuchsian equations with smooth 
coefficients in the literature and to start with we need to understand 
why these do not apply directly to Gowdy spacetimes. The general form of 
a system of Fuchsian equations for a vector-valued unknown function $u$ is 
\begin{equation}\label{fuchs}
t\frac{\d u}{\d t}+N(x)u=tf(t,x,u,D_x u)
\end{equation}
Here $x$ is a point in some Euclidean space and $D_x u$ is a shorthand for 
the first order derivatives of $u$ with respect to the spatial variables $x$.
The function $f$ is required to extend continuously to $t=0$ while the 
matrix $N(x)$ is required to satisfy some positivity condition, which may 
depend on the particular context. An example of a condition of this type 
is that the eigenvalues of the matrix $N(x)$ should have non-negative real 
parts for all $x$. Note that the apparently more general system
\begin{equation}\label{fuchsalpha}
t\frac{\d u}{\d t}+N(x)u=t^\alpha f(t,x,u,D_x u)
\end{equation}
with $\alpha>0$ can be reduced to the form (\ref{fuchs})
by introducing $t^\alpha$ as a new time variable. This results in the
matrix $N(x)$ being rescaled by a factor $\alpha$, but this does not
affect its positivity properties.

One approach to proving existence theorems for Fuchsian systems which does
not require any analyticity assumption is due to Claudel and Newman
\cite{claudel98}. Of course, in the context of data which are merely smooth
(or even finitely differentiable) the system must be hyperbolic. This is
needed to prove existence for an equation without any singular behaviour
in $t$ and the singularity cannot be expected to improve the situation.
In the Claudel-Newman theorem it is required that $f$ have an asymptotic
expansion about $t=0$ 
in integral powers of $t$ and this property is inherited by the solution.
The positivity assumption on $N(x)$ is weaker than that mentioned
above. It is only required that there be no eigenvalues which are negative
integers. An important element of the proof of the theorem is to expand the
candidate solution $u$ in powers of $t$, writing it in the form $u_0+u_1$
where $u_0$ is an appropriate polynomial in $t$ and $u_1$ is a remainder of
higher order in $t$. Then $u_1$ solves a Fuchsian system where the 
eigenvalues of $N(x)$ are shifted by an integer in the positive direction.
If it is possible to expand to a sufficiently high order then the shifted
eigenvalues all have positive real parts. The condition which makes the
expansion possible is that a polynomial $u_0$ can be found which satisfies
the equation up to an error of sufficiently high order in $t$. The only
obstruction to this is if the shifted eigenvalue becomes zero at some
stage in the process and this is prevented by the assumption made on the
eigenvalues of the original matrix. In cases such as the system arising
in the analysis of Gowdy spacetimes in \cite{kichenassamy98} the solutions 
cannot be expanded in integral powers of $t$. Instead non-integral (and even
$x$-dependent) powers of $t$ and logarithms occur. For this reason the
method of \cite{claudel98} does not apply directly. One of the main methods
of the present paper is to extend the technique of using expansions of the
solution to shift the eigenvalues of $N(x)$ to cases where terms more
complicated than integral powers of $t$ occur.

Another approach to proving existence theorems for Fuchsian systems with
smooth coefficients is due to Kichenassamy\cite{kichenassamy96a}. In that 
case there is no restriction that solutions have expansions in 
integral powers of $t$. It is, however, required that the matrix $N(x)$ be 
independent of $x$. This is not satisfied in the Gowdy case. Since the 
eigenvalues of the matrix $N(x)$ correspond to powers occurring in the 
expansions, this is a consequence of the dependence of these powers on $x$ 
in the system coming from the Gowdy spacetimes. Thus the result of 
\cite{kichenassamy96a} does not apply. It might be possible to extend the 
proofs in \cite{kichenassamy96a} to the case of non-constant $N$, but this 
will not be attempted here. Both the proofs of Claudel and Newman and of 
Kichenassamy involve the use of sophisticated techniques from functional 
analysis, namely semigroup theory and the Yosida approximation, 
respectively. These will be avoided in the approach developed in the 
following. 

In the paper \cite{anguige99d} Anguige treats the case of plane symmetric 
solutions of the Einstein equations coupled to a perfect fluid. He uses a 
direct energy argument of a type familiar in the theory of regular 
symmetric hyperbolic systems. It is important in his proof that $N(x)$ is 
independent of $x$ and positive semi-definite. This kind of argument will be 
generalized in the following. On the one hand the condition that $N(x)$ 
should be constant will be removed. On the other hand the positivity 
condition on the matrix will be relaxed by a method related to that used 
in \cite{claudel98}. The eigenvalue condition used in \cite{claudel98} will 
be replaced by a condition of formal solvability which abstracts its 
essential significance.

Next a brief survey of the literature on applications of Fuchsian equations 
to general relativity will be given. It appears that the first paper
containing an application of this kind was \cite{moncrief82},
where Moncrief proved the existence of a large class of analytic spacetimes
with analytic compact Cauchy horizons. Later Newman 
\cite{newman93a, newman93b} 
based his work on isotropic singularities and Penrose's Weyl curvature 
hypothesis on existence theorems for hyperbolic systems with singularities 
of Fuchsian type. These papers did not include proofs of the 
required theorems but the necessary proofs were provided in \cite{claudel98}. 
More recently, results on isotropic singularities for more general matter 
models were obtained by Anguige and Tod \cite{anguige99a, anguige99b, 
anguige99c}. Their theorems 
require no symmetry assumptions but are confined to a special type of 
singularities. On the other hand Anguige\cite{anguige99d} proved a theorem
on the existence of non-isotropic singularities in plane symmetric spacetimes
with perfect fluid.

Another line of development of the applications of Fuchsian equations in
general relativity starts with the work of Kichenassamy and Rendall
\cite{kichenassamy98} on singularities in analytic Gowdy spacetimes.
It builds on previous work of Kichenassamy  outside general relativity  
(see \cite{kichenassamy96b}
and references therein). This direction is continued in the papers 
\cite{isenberg99}, \cite{andersson00} and \cite{narita00} which deal with 
vacuum models with two
spacelike Killing vectors, solutions of the Einstein-scalar field
equations and analogues of Gowdy models in string cosmology, respectively.
In all these cases analyticity is assumed. A notable feature of the result 
of \cite{andersson00} is that it makes no symmetry assumptions and so, on
the basis of function-counting arguments, concerns general solutions of the
Einstein equations with the given matter model.  

The paper is organized as follows. In the second section the notion of a 
formal solution of a Fuchsian system is defined. Assumptions on the 
coefficients are exhibited which guarantee
the existence of a formal solution. They are fulfilled by the first order
form of the Gowdy equations introduced in \cite{kichenassamy98}. In the third
section this form of the equations is modified slightly so as to obtain a
symmetric hyperbolic system. Its formal solvability is shown to follow from
that of the original system. The system satisfied by the remainder term 
which is the difference between a true solution of the system and an
approximate solution is computed. The fourth section proves an existence 
theorem which is general enough to apply to the case of Gowdy spacetimes
with sufficiently low velocity ($k<3/4$). In the fifth section yet another 
form of the equations is used to cover the remainder of the full low velocity 
case ($k<1$). The wider applicablity of the methods of the paper is discussed 
in the final section.

\section{Formal solutions}\label{formal}

If the function $f$ in (\ref{fuchs}) is smooth at $t=0$ and hence admits an
asymptotic expansion about $t=0$ in integral powers of $t$ then it can be
useful to expand it in this way. In the following a generalization to less
smooth functions $f$ will play an important role. This uses the notion of
a formal solution of equation (\ref{fuchs}) which will now be defined. It 
will also be convenient for the following to introduce a notion of regularity
of functions adapted to the given situation. An analogous notion in an 
analytic context was introduced in \cite{andersson00}.

\noindent
{\bf Definition 1} A function $z(t,x)$ defined on an open subset of 
$[0,\infty)\times R^N$ and taking values in $\R^K$ is called {\it regular}
if it is $C^\infty$ for all $t>0$ and if its partial derivatives (defined 
for $t>0$) of any order with respect to the variables $x\in\R^K$ extend 
continuously to $t=0$.  

\noindent
{\bf Definition 2} A finite sequence $(u_1,u_2,\ldots,u_p)$ of functions 
defined on an open subset $U$ of $[0,\infty)\times\R^n$ containing 
$\{0\}\times\R^n$ is called a {\it formal solution} of order $p$ of 
(\ref{fuchs}) if
\begin{enumerate}
\item each $u_i$ is regular
\item $t\d_t u_i+N(x)u_i-tf(t,x,u_i,D_x u_i)=O(t^i)$ for all $i$ as 
$t\to 0$
\end{enumerate}
Here and in the following the $O$-symbols are taken in the sense of uniform
convergence on compact subsets. 

In \cite{kichenassamy98} an iteration was defined which, in the case that the
function $f$ has suitable analyticity properties, converges to a solution of
(\ref{fuchs}). It is doubtful if it converges in any useful sense when $f$
is only smooth. However it can be used to produce a formal solution of any
desired order, as will now be shown.

\noindent
{\bf Lemma 2.1} If the function $f$ is regular and the matrix $N(x)$ is 
smooth and satisfies an estimate of the form $\|\sigma^{N(x)})\|\le C$ with 
a constant $C$ for $\sigma$ in a neighbourhood of zero then for each $p$
the equation (\ref{fuchs}) has a formal solution of order $p$ which 
vanishes at $t=0$.

\noindent
{\bf Proof} First some definitions from \cite{kichenassamy98} will be 
recalled. For a regular function $u$ define $F[u]=tf(t,x,u,D_x u)$. Then
$F[u]$ is also regular and is $O(t)$ as $t\to 0$ together with all its
spatial derivatives. If $v$ is regular and $O(t)$ as $t\to 0$ together with
all its spatial derivatives, define 
$u=H[v]=\int_0^1 \sigma^{N(x)-I}v(\sigma t)d\sigma$. Then $u$ is
regular and $O(t)$ together with all its spatial derivatives and satisfies 
$(t\d_t+N)u=v$. Then, if $G$ is defined to be the composition $FH$, any 
fixed point $v$ of $G$ defines a solution $u$ of (\ref{fuchs}) by $u=H[v]$. 
Let $u_1=0$. It defines a formal solution of (\ref{fuchs}) of order one. 
It will be shown that defining $u_i=HG^{i-1}[u_1]$ defines a formal 
solution of order $p$ for each $p$. Note the relation $u_{i+1}=HFu_i$.

The first defining property of a formal solution is easily proved
by induction. The main point is to verify the second property. To do this 
it will first be shown that $u_{i+1}-u_i=O(t^i)$ for each $i$, and that 
spatial derivatives of all orders of the $u_i$ satisfy the corresponding 
estimates. For $i=1$ this follows directly from the properties already
demonstrated. To prove the general case, consider the equation obtained
by forming the difference of the equations satisfied by $u_{i+1}$ and 
$u_i$. This gives
\begin{equation}\label{udiff}
t\d_t (u_{i+1}-u_i)+N(x)(u_{i+1}-u_i)=tM_1(u_i-u_{i-1})+tM_2 D_x(u_i-u_{i-1})
\end{equation}
for regular functions $M_1$ and $M_2$ of the arguments $t$, $x$, $u_i$,
$u_{i-1}$, $D_x u_i$, $D_x u_{i-1}$, obtained by applying the mean value 
theorem
to differences. The right hand side of (\ref{udiff}) is $O(t^i)$. Then
the fact can be applied that the operator $H$ preserves the set of functions
which are $O(t^j)$ for any $j$. Spatial derivatives can be handled in the
same way. To complete the proof of the lemma, consider the relation:
\begin{eqnarray}\label{remainder}
&&t\d_t u_{i+1}+N(x)u_{i+1}-tf(t,x,u_{i+1},D_x u_{i+1})\nonumber        \\
&&=-t[f(t,x,u_{i+1},D_x u_{i+1})-f(t,x,u_i,D_x u_i)]
\end{eqnarray} 
Using the mean value theorem and the estimates already obtained for 
$u_{i+1}-u_i$ shows that the right hand side of (\ref{remainder}) is 
$O(t^{i+1})$.

\noindent
{\bf Remark} A general criterion for checking the condition on $N(x)$ required
to apply this lemma has been given in (\cite{andersson00}).
  
Next some basic equations for the Gowdy spacetimes will be recalled. More
details can be found in \cite{kichenassamy98}. The basic unknowns are two
real-valued functions $X(t,x)$ and $Z(t,x)$ of two variables. New variables
$u$ and $v$ are defined so that the relations
\be\label{zansatz}
Z(t,x)=k(x)\log t+\phi(x)+t^\epsilon u(t,x)
\ee      
and
\be\label{xansatz}
X(t,x)=X_0(x)+t^{2k(x)} (\psi(x)+v(t,x))
\ee
hold, where $k$, $X_0$, $\phi$ and $\psi$ are given functions. The positive 
constant $\epsilon$ will be restricted later. Next introduce further 
variables by setting 
\be
(u_0,u_1,u_2,v_0,v_1,v_2)=(u,t\d_t u,tu_x,v,t\d_t v,tv_x)
\ee
The Gowdy equations imply the following first order system:
\bea\label{reduced1}
t\d_t u_0&=&u_1             \\
t\d_t u_1&=&-2\epsilon u_1-\epsilon^2 u_0+t^{2-\epsilon}(k_{xx}\log t
+\phi_{xx})
+t\d_x u_2\nonumber             \\ 
&&-\exp(-2\phi-2t^\epsilon u_0)\{t^{2k-\epsilon}(v_1+2kv_0+2k\psi)^2 
-t^{2-2k-\epsilon}X_{0x}^2\nonumber   \\
&&-2t^{1-\epsilon}X_{0x}(v_2+t\psi_x+k_x(v_0+\psi)t\log t) \nonumber \\
&&-t^{2k-\epsilon}(v_2+t\psi_x+2k_x (v_0+\psi) t\log t)^2\}         \\
t\d_t u_2&=&t\d_x(u_0+u_1)                   \\
t\d_t v_0&=&v_1             \\       
t\d_t v_1&=&-2kv_1+t^{2-2k}X_{0xx}+t\d_x(v_2+t\psi_x)+4k_x(v_2+t\psi_x)t\log t 
\nonumber              \\
&&+(v_0+\psi)[2k_{xx}t^2\log t+4(k_x t\log t)^2]\nonumber   \\
&&+2t^\epsilon (v_1+2kv_0+2k\psi)(u_1+\epsilon u_0)\nonumber  \\
&&-2X_{0x}t^{2-2k}(k_x\log t+\phi_x+t^\epsilon \d_x u_0)\nonumber \\
&&-2t(\d_x(v_0+\psi)+2k_x(v_0+\psi)\log t)(k_xt\log t+t\phi_x+t^\epsilon u_2)
\\ 
t\d_t v_2&=&t\d_x(v_0+v_1)
\eea
Here some minor errors in the equations given in \cite{kichenassamy98} have
been corrected\footnote{I thank Aurore Cabet for pointing these out}. 
This system is of the form (\ref{fuchsalpha}) which implies, 
as indicated in the introduction, that it can be brought into the form 
(\ref{fuchs}) by a change of time variable. This possibility  will be used 
freely without further comment in the following. After the change of time 
coordinate the system arising from the Gowdy equations satisfies the 
hypotheses of Lemma 2.1 provided $\epsilon<2k$ and $\epsilon<2-2k$. In 
particular, the bound on $\sigma^{N(x)}$ was verified directly in 
\cite{kichenassamy98}. Alternatively, it follows easily from the criterion
given in \cite{andersson00}. Hence the above system has a formal solution 
of any order which vanishes at $t=0$.

\section{The symmetric hyperbolic system}\label{symhyp}

In the following the form of the Gowdy equations introduced in section
\ref{formal} will be called the first reduced system. Now it will be 
modified to get a system which, while less convenient for showing formal
solvability, is symmetric hyperbolic and therefore appropriate for allowing
the theory of hyperbolic equations to be applied. It is obtained
from the first reduced system by making the substitutions $u_2=t\d_x u_0$
and $v_2=t\d_x v_0$ in some places. The result is:
\bea\label{reduced2}
t\d_t u_0&=&u_1             \\
t\d_t u_1&=&-2\epsilon u_1-\epsilon^2 u_0+t^{2-\epsilon}(k_{xx}\log t
+\phi_{xx})
+t\d_x u_2\nonumber             \\ 
&&-\exp(-2\phi-2t^\epsilon u_0)\{t^{2k-\epsilon}(v_1+2kv_0+2k\psi)^2 
-t^{2-2k-\epsilon}X_{0x}^2\nonumber   \\
&&-2t^{1-\epsilon}X_{0x}(v_2+t\psi_x+k_x(v_0+\psi)t\log t) \nonumber \\
&&-t^{2k-\epsilon}(v_2+t\psi_x+2k_x (v_0+\psi) t\log t)^2\}         \\
t\d_t u_2&=&u_2+t\d_x u_1                   \\
t\d_t v_0&=&v_1             \\       
t\d_t v_1&=&-2kv_1+t^{2-2k}X_{0xx}+t\d_x(v_2+t\psi_x)+4k_x(v_2+t\psi_x)t\log t 
\nonumber              \\
&&+(v_0+\psi)[2k_{xx}t^2\log t+4(k_x t\log t)^2]\nonumber   \\
&&+2t^\epsilon (v_1+2kv_0+2k\psi)(u_1+\epsilon u_0)\nonumber  \\
&&-2X_{0x}t^{2-2k}(k_x\log t+\phi_x+t^{\epsilon-1}u_2)\nonumber \\
&&-2(v_2+t\d_x\psi+2tk_x(v_0+\psi)\log t)(k_xt\log t+t\phi_x+t^\epsilon u_2)
\\ 
t\d_t v_2&=&v_2+t\d_x v_1
\eea
This system, which will be referred to as the second reduced system, has the
advantage of being symmetric hyperbolic but also has two potential 
disadvantages. The one is that the matrix $N(x)$ has been modified while the
second is that possibly dangerous powers of $t$ have been introduced on the
right hand side. The matrix $N(x)$ acquires two negative eigenvalues, which
will have to be dealt with by appropriate methods in due course. As far as 
the other problem is concerned, the power $t^{1+\epsilon-2k}$ is introduced.
This power should be positive. This can be achieved subject to the 
inequalities already assumed for $\epsilon$ if and only if $k<3/4$. This 
restriction appears unnatural, but will be assumed in this section and the
next for the second reduced system.

In fact, although the desired inequalities relating $\epsilon$ and $k(x)$ 
can be ensured by a suitable choice of $\epsilon$ at any given point $x$,
this cannot be done at all points simultaneously by a single choice of the
constant $\epsilon$. It can be ensured in a neighbourhood of any given point.
Solutions will be constructed in local neighbourhoods of this kind and then
put together using the domain of dependence to get a solution which is global
in $x$.

Any formal solution $\{u_1,\ldots,u_p\}$ of the first reduced system which
vanishes at $t=0$ is also a formal solution of the second reduced system. 
For $t\d_t((u_2-t\d_x u_0)_i)=O(t^i)$ and using the fact that the formal
solution vanishes at $t=0$ it follows that
$(u_2-t\d_x u_0)_i=O(t^i)$. This can then be used to see that the
difference terms arising when passing from the first to the second 
reduced systems are $O(t^i)$, assuming the condition $k<3/4$.
Although it is not of significance for the following it is interesting
to note that for $3/4<k<1$ the sequence whose element with index $i$
is the element of the formal solution of the first reduced system with
index $i+1$ is a formal solution of the second reduced system.

Given a formal solution it is possible to consider the difference between
an actual solution and the formal solution and the equations which this
difference satisfies. The hope is that this equation is more tractable
analytically than the first and second reduced systems. This is a 
generalization of the procedure of subtracting a Taylor polynomial of
finite order in the case that the solutions are smooth at $t=0$. Rather
than doing this calculation in the specific case of the Gowdy system it
will be done for the following more general symmetric hyperbolic Fuchsian
system:
\begin{equation}\label{symhypeq}
t\d_t u+N(x)u+tA^j(t,x,u)\d_j u=tf(t,x,u)
\end{equation} 
The condition for symmetric hyberbolicity is that the matrices $A^j$ should
be symmetric. As before, all coefficients in the equation are assumed 
regular. If $\{u_i\}$ is a formal solution and $v_i=t^{-i}(u-u_i)$ then
\begin{equation}\label{diffeq}
t\d_t v_i+(N(x)+iI)v_i+tA^j(t,x,u_i+t^iv_i)\d_j v_i=tg_i(t,x,v_i)
\end{equation}
for some regular function $g_i$. Choosing $i$ large enough
ensures that the eigenvalues of $N+iI$ have positive real parts, or even
that the matrix is positive definite. In the Gowdy case this will be referred
to as the third reduced system.

\section{The basic existence theorem}\label{existence}

In this section a local existence theorem will be proved for solutions of 
the Gowdy equations in a neighbourhood of the initial singularity in the 
case that the data $k$, $X_0$, $\phi$ and $\psi$ are merely smooth. In the 
case of analytic data the problem was solved in \cite{kichenassamy98}. Thus 
if the data are approximated by a sequence of analytic data $(k_m,X_{0m},
\phi_m,\psi_m)$, a corresponding sequence of analytic solutions is obtained.
At the same time formal solutions can be obtained for both the 
approximate data and the actual data. Denote the former by $u_{mi}$, where
the first index corresponds to the sequence of data and the second to the
enumeration of elements of an approximate solution. Denote the latter by
$u_i$, as before. If the approximate solutions are constructed as in
the proof of the Lemma 2.1 then $u_{mi}\to u_i$ as $m\to\infty$,
uniformly on compact subsets. The same is true of the spatial derivatives
of these functions of any order. It can be concluded that the sequence of
coefficients obtained for the third reduced system is also convergent on
compact sets as $m\to\infty$. The sequence of solutions of these equations
which we have is only defined a priori on a time interval which depends on
$m$. However, using the global existence theorem for the Gowdy equations
\cite{moncrief80} it is possible to conclude that a sequence of smooth 
solutions exists on a common time interval. The aim now is to show that 
this is a Cauchy sequence in a sufficiently strong topology. If that can
be done then it will follow that the sequence converges to a limit which 
is a solution corresponding to the smooth data originally prescribed.

The tool to obtain convergence of the approximations is the technique
of energy estimates. This requires some preliminary remarks on linear
algebra. Consider the matrix-valued function $N(x)$ in the Fuchsian 
system. Spatial derivatives of a solution of this system also satisfy a system
of the same form, but with a different matrix $N(x)$. Suppose, for instance,
we consider the first derivative $D_xu$ of the unknown. The system for the
pair $(u,D_xu)$ has a matrix in its singular term which has diagonal blocks
$N(x)$ and an off-diagonal block involving $D_xN(x)$. This off-diagonal 
block does not affect the eigenvalues of the matrix but may well affect 
whether it is positive definite or not. Since the positive definiteness
of matrices like this is important in what follows we adopt a strategy
which avoids positivity being lost. In order to implement this it will be
assumed that $N(x)$ is positive definite. Then use the variable $w=KD_xu$ 
for a positive constant $K$ instead of $D_x u$ itself. For the equation 
satisfied by $(u,w)$ the matrix of interest is positive definite provided 
$K$ is chosen sufficiently small. The same trick works for higher 
derivatives. It suffices to replace the collection of unknowns 
$\{D^\alpha u\}$ by $w^\alpha=K^{|\alpha|}D^\alpha u$. Let the
matrix corresponding to $N$ arising in the system for all these derivatives
up to order $s$ be denoted by $N^{(s)}$. By construction it is positive 
definite.

The standard method of energy estimates (see e.g. \cite{taylor96} or, for a
discussion aimed at relativists, \cite{friedrich00}) proceeds
by estimating the Sobolev norms of solutions. The usual Sobolev norm is
given by $\|u\|_{H^s}=(\sum_{|\alpha |\le s}\|D^\alpha u\|^2_{L^2})^{1/2}$.
For the present purposes it is convenient to use the equivalent norm
$\|u\|_{H^s, K}=(\sum_{|\alpha |\le s}K^{2|\alpha|}
\|D^\alpha u\|^2_{L^2})^{1/2}$, where $K$ is, as before, a small enough 
positive constant.

As a first application of energy estimates, a theorem on the domain of
dependence will be proved. Let $u$ and $v$ be two regular solutions of
(\ref{symhypeq}) vanishing at $t=0$. Then their difference satisfies an 
equation of the following form:
\begin{equation}\label{difference}
\d_t (u-v)+t^{-1}N(x)(u-v)+A^j(t,x,u)\d_j (u-v)=M(t,x) (u-v)
\end{equation}  
where $M$ is a regular function constructed from $u$ and $v$. Choose two
times $t_1$ and $t_2$ with $0<t_1<t_2$ and let $G$ be the region defined
by the inequalities $t_1\le t\le t_2$ and $|x|\le 2t_2-t$. Let $S_1$ and
$S_2$ be its intersections with $t=t_1$ and $t=t_2$ respectively. Now 
multiply the equation (\ref{difference}) by $e^{-\kappa t}(u-v)$, integrate 
over $G$ and integrate by parts in the way this is usually done in the 
derivation of energy estimates. Here $\kappa$ is a positive constant.
The singular term containing $N$ can be discarded, due to its sign, giving
an estimate of the form $e^{-\kappa t_2}I_2\le e^{-\kappa t_1}I_1+I_G$, 
where $I_1$ and $I_2$ are the $L^2$ norms of the restrictions of $u-v$ to 
$S_1$ and $S_2$ respectively, and $I_G$ is a volume integral over $G$ which
for $\kappa$ sufficiently large is negative unless $u-v$ is identically 
zero on $G$. Letting $t_1$ tend to zero, so that $I_1\to 0$, gives a 
contradiction unless $u-v=0$ on $G$. Thus it has been proved that the 
solutions $u$ and $v$ agree on $G$. This proves a domain of dependence 
property for solutions of (\ref{symhypeq}) which can be used for the 
purpose of gluing together solutions. This means that even if we are 
interested in producing solutions on manifolds it is enough to solve the 
problem on $\R^n$. Moreover, it is possible to consider without loss of 
generality the case of cut-off coefficients and data. By this we mean that 
there is a compact subset of $\R^n$ such that for $x$ outside this compact 
set the initial data vanish, the coefficients $A^j$ and $f$ vanish and $N$ 
is constant.

The aim is now to construct solutions of the third reduced system for Gowdy
in the case of smooth coefficients. As already indicated it is enough to do
this under the assumption of a cut-off in space. A sequence of functions
which is a candidate for a sequence converging to a solution of the third
reduced system has already been produced. This sequence is obtained by
fixing a value of $i$ sufficiently large that the matrix occurring in the 
singular term of the third reduced system is positive definite and forming 
the difference of the solution of the second reduced system corresponding
to the data $(k_m,\phi_m,X_{0m},\psi_m)$ and the function $u_{mi}$. 
It will now be shown by using
energy estimates that this sequence is bounded in suitable Sobolev spaces
and in fact is a Cauchy sequence. To do this differentiate 
$\|u\|_{H^s,K}^2$ with respect to $t$ and substitute the third reduced
system into the result. This gives
\begin{equation}\label{energy}
d/dt(\|u\|_{H^s,K})^2)=-t^{-1}\langle N^{(s)}u^{(s)},u^{(s)}
\rangle_{L^2,K}+R
\end{equation}
Here $u^{(s)}$ is the collection of all derivatives of $u$ up order $s$ and
$R$ is the sum of the terms which arise in the regular case, i.e. in the
case where $N$ is identically zero. The first term on the right hand side
is non-positive and may
be discarded. The terms in $R$ can be estimated just as in the regular case
and this gives a bound for the $H^s$ norm of $u$, provided $s>n/2+1$.
Next the difference of successive approximants will be estimated. An attempt 
to apply the standard techniques in order to show that the sequence is Cauchy
only meets one difficulty not present in the regular Cauchy problem. This
is due to differences of the matrices $N$ for successive elements of the 
sequence. This can be overcome in a way similar to that used above where
the norms were scaled. The trick is to consider the collection of the 
derivatives of the functions of the sequence up to order $s$ together with
the derivatives of the differences of successive elements of the sequence
up to order $s-1$, the derivatives of the differences being multiplied 
by an additional factor $K$. This once again ensures that the final matrix
obtained is positive definite.

It follows from the above discussions that the sequence of solutions $v_i$
of the third reduced system converges to a solution of the third reduced 
system corresponding to the original smooth data. This can then be used
to define solutions of the second and first reduced systems and finally
a solution of the Gowdy system itself corresponding to the data we started
with. The construction is such that the interval on which convergence is
obtained may depend on $s$. However standard results about symmetric 
hyperbolic systems with smooth coefficients show that the solutions
for all values of $s$ can be extended to a common time interval. Hence 
a solution is obtained on that time interval which is $C^\infty$
for $t>0$. The results of this discussion are summed up in the following 
theorem, which may be compared with Theorem 1 of \cite{kichenassamy96a}.

\noindent
{\bf Theorem 4.1} Let $k(x)$, $X_0(x)$, $\phi(x)$ and $\psi(x)$ be $C^\infty$
and assume that $0<k(x)<3/4$ for all $x$. Then there exists a solution of the 
Gowdy equations with following properties. For each spatial point $x$ there
exists an open neighbourhood $U_x$ of $x$ and a number $\epsilon_x>0$ such
that the restriction of the solution to $U_x$ satisfies (\ref{zansatz}) and 
(\ref{xansatz}) with $\epsilon=\epsilon_x$, where $u$ and $v$ are regular 
and tend to zero as $t\to 0$. The $U_x$ and $\epsilon_x$ can be chosen in
such a way that the inequalities $2k(y)-1<\epsilon_x<\min\{2k(y),2-2k(y)\}$ 
are satisfied for all $x$ and all $y\in U_x$. Under these conditions the 
solution is unique. 

\noindent
{\bf Remark} A formulation of the theorem which is equivalent but cleaner
can be obtained by replacing the constant $\epsilon$ by a function 
$\epsilon(x)$. Then it would not be necessary to introduce the $U_x$.

In \cite{kichenassamy96a} a theorem was proved concerning high velocity
analytic solutions in the case where $X_0$ is independent of $x$. The 
method used to prove Theorem 4.1 applies straightforwardly to the high
velocity case to give an analogue of the result of \cite{kichenassamy96a}
in the smooth case. The following theorem results.

\noindent
{\bf Theorem 4.2} Let $k(x)$, $\phi(x)$ and $\psi(x)$ be $C^\infty$
and let $X_0$ be a constant. Assume that $k(x)>0$. Then there exists a 
solution of the Gowdy equations with the following properties. For each 
spatial point $x$ there exists an open neighbourhood $U_x$ of $x$ and a 
number $\epsilon_x>0$ such that the restriction of the solution to $U_x$ 
satisfies (\ref{zansatz}) and (\ref{xansatz}) with $\epsilon=\epsilon_x$, 
where $u$ and $v$ are regular and tend to zero as $t\to 0$. The $U_x$ and 
$\epsilon_x$ can be chosen in such a way that the inequality $\epsilon_x
<2k(y)$ is satisfied for all $x$ and all $y\in U_x$. Under these conditions 
the solution is unique.

\section{Data with intermediate velocity}

In the previous section an existence theorem was proved for Gowdy spacetimes
under the restriction $0<k<3/4$ on the function $k$. Next it will be shown
that using a different ansatz allows the range $1/2<k<1$ to be treated. The
two together then cover the whole range $0<k<1$ for which results were 
available in the analytic case. The new ansatz involves expanding the 
function $Z$ to a higher order in $t$. The ansatz for $X$ remains unchanged.
Now $Z$ is of the form
\begin{equation}\label{zansatz2}
Z=k\log t+\phi+\alpha t^{2-2k}+t^{2-2k+\epsilon}u
\end{equation} 
where $\alpha=(2-2k)^{-2}X_{0x}^2$.
Reexpressing the Gowdy equations in terms of the new variables $u$ and $v$
and reducing to first order as before leads to an analogue of
the second reduced system of section \ref{symhyp}. The evolution equations for
$u_0$, $u_2$, $v_0$ and $v_2$ are the same as before and will not be repeated.
The modified equation for $u_1$ is
\bea\label{u1new}
t\d_t u_1&=&-2(2-2k+\epsilon) u_1-(2-2k+\epsilon)^2 u_0
+t^{2k-\epsilon}(k_{xx}\log t+\phi_{xx})
+t\d_x u_2\nonumber             \\ 
&&+t^{1-\epsilon}[t\alpha_{xx}-4k_x\log t(t\alpha_x+t^\epsilon u_2)
+4k_x^2t(\log t)^2(\alpha+t^\epsilon u_0)]\nonumber           \\
&&-\exp(-2\phi-2\alpha t^{2-2k}-2t^{2-2k+\epsilon} u_0)
\{t^{4k-2-\epsilon}(v_1+2kv_0+2k\psi)^2\nonumber   \\ 
&&-2t^{2k-1-\epsilon}X_{0x}(v_2+t\psi_x+k_x(v_0+\psi)t\log t) \nonumber \\
&&-t^{4k-2-\epsilon}(v_2+t\psi_x+2k_x (v_0+\psi) t\log t)^2\}
\eea
and that for $v_1$ is
\bea\label{v1new}
t\d_t v_1&=&-2kv_1+t^{2-2k}X_{0xx}+t\d_x(v_2+t\psi_x)+4k_x(v_2+t\psi_x)t\log t 
\nonumber              \\
&&+(v_0+\psi)[2k_{xx}t^2\log t+4(k_x t\log t)^2]\nonumber   \\
&&+2t^{2-2k+\epsilon} (v_1+2kv_0+2k\psi)(u_1+(2-2k+\epsilon) u_0)\nonumber \\
&&+(4-4k)t^{2-2k}(v_1+2kv_0+2k\psi)\alpha  \\
&&-2X_{0x}t^{2-2k}(k_x\log t+\phi_x+t^{2-2k}\alpha_x+t^{1-2k+\epsilon}u_2
\nonumber \\
&&-2k_xt^{2-2k}\log t(\alpha+t^\epsilon u_0))
-2(v_2+t\d_x\psi+2tk_x(v_0+\psi)\log t)\times\nonumber   \\
&& (k_xt\log t+t\phi_x+t^{3-2k}\alpha_x
+t^{2-2k+\epsilon} u_2-2k_x t^{3-2k}\log t
(\alpha+t^{\epsilon}u_0)\nonumber
\eea
There is also an obvious analogue of the first reduced system of section
\ref{formal}. The existence of formal solutions of the latter is
guaranteed by Lemma 2.1 and these give rise to formal solutions of
the second reduced system as in section \ref{symhyp}.

Consider now the sequence of analytic solutions of the Gowdy equations
corresponding to a sequence of analytic approximations to the smooth
data of interest. It will be shown that, under the condition that
$1/2<k<1$, these define a sequence of regular solutions of the second 
reduced system of this section. To do this it is necessary to show that 
for each of these solutions the function $Z$ admits an asymptotic 
expansion of the form (\ref{zansatz2}) and not just of the form 
(\ref{zansatz}), which is known a priori. To do this it suffices to 
apply the existence theorem of \cite{kichenassamy98} to the first 
reduced system of this section with the analytic data.

Once these facts are known, it is straightforward to prove an existence 
theorem for the second reduced system of this section using the same 
techniques as were applied to the second reduced system of section 
\ref{symhyp} provided certain inequalities are satisfied. These are the 
inequalities which ensure that each term on the right hand side of
the equation contains a positive power of $t$. Under the assumption that
$1/2<k<1$ this can be achieved by choosing the positive real number 
$\epsilon$ to satisfy $4k-3<\epsilon<2k-1$. The following theorem is
obtained.

\noindent
{\bf Theorem 5.1} Let $k(x)$, $X_0(x)$, $\phi(x)$ and $\psi(x)$ be $C^\infty$
and assume that $1/2<k(x)<1$ for all $x$. Then there exists a solution of the 
Gowdy equations with the following properties. For each spatial point $x$ 
there exists an open neighbourhood $U_x$ of $x$ and a number $\epsilon_x>0$ 
such that the restriction of the solution to $U_x$ satisfies (\ref{zansatz2}) 
and (\ref{xansatz}) with $\epsilon=\epsilon_x$, where $u$ and $v$ are regular 
and tend to zero as $t\to 0$. The $U_x$ and $\epsilon_x$ can be chosen in 
such a way that the inequalities $4k(y)-3<\epsilon_x<2k(y)-1$ are satisfied 
for all $x$ and all $y\in U_x$. Under these conditions the solution is unique.

\section{Discussion}\label{discussion}

The theorems stated in this paper have all concerned Gowdy spacetimes. It
is nevertheless clear that many of the arguments are much more generally 
applicable. At the same time some steps are essentially related to the 
specific Gowdy case. A general discussion of the procedure will now be given
which separates the general from the particular as much as possible. The 
first step is to make a suitable ansatz for the solutions to be constructed
as the sum of an explicit part and a remainder. There may be more than one 
useful way of doing this. For example, in the Gowdy case equations 
(\ref{zansatz}) and (\ref{xansatz}) were useful for proving one theorem
while replacing (\ref{zansatz}) by (\ref{zansatz2}) allowed a different
theorem to be proved. The second step is to reduce the equations to first
order. The aim is to produce a system of Fuchsian form for which the 
theorem of \cite{kichenassamy96a} ensures the existence of 
solutions corresponding to the case where the free functions in the 
ansatz are analytic. If these free functions are merely smooth the lemma 
proved in section 2 may be used to show the existence of formal solutions.

The third step is to produce a system which is symmetric hyperbolic and
in Fuchsian form. At this stage the matrix $N(x)$ may have negative 
eigenvalues, as is the case in the Gowdy example. It needs to be shown
that the formal solutions already produced define formal solutions of
the symmetric hyperbolic system. From this point on the argument proceeds
on a general level, with no more details of the Gowdy special case being
used.

It is instructive at this stage to consider what difficulties would be
likely to arise in an attempt to generalize the results of \cite{andersson00}
from the analytic to the smooth case. One problem is to bring the equations
into a suitable hyperbolic form by the choice of coordinate or
gauge conditions. There was no difficulty of this type in the Gowdy case,
where a rather rigid preferred coordinate system is available. In more
general cases it will be necessary to choose a form of the reduced Einstein
equations carefully from the myriad on offer. If a symmetric hyperbolic 
system is obtained it is likely to involve a matrix $A^0$ multiplying the
time derivative of the unknown which is not the identity, thus going beyond
the case discussed above. Even worse, it may be difficult to ensure that
$A^0$ remains bounded and uniformly positive definite as $t\to 0$. These
conditions are very important for the use of energy estimates.

To conclude the paper, an application will be presented where the 
flexibility of smooth functions is essential. Existence theorems 
have been proved for Gowdy spacetimes in the low velocity case and,
under the condition that $X_0$ is constant, also in the high velocity
case. Using the domain of dependence these can be combined to give a
more general class of solutions. To do this consider a smooth function
$X_0$ which is constant on a non-empty open interval $I$. Now complete
this to data $(k,X_0,\phi,\psi)$ in such a way that $k<1$ on the closure 
of the complement of $I$. Then each point $x$ has a neighbourhood on 
which one of the 
existence theorems applies and the resulting local solutions can be put 
together to produce a solution corresponding to the chosen initial data 
globally in $x$. If we tried to do this construction with analytic data 
then $X_0$ would have to be globally constant and nothing new would be 
obtained.

\end{document}